# Emergence of Superconductivity in Doped Multiorbital Hubbard Chains

Niravkumar D. Patel[1], Nitin Kaushal[2,3], Alberto Nocera[2,4], Gonzalo Alvarez[5], Elbio Dagotto[2,3,*]

1 Department of Physics, The Ohio State University, 191 W. Woodruff Avenue, Columbus, OH 43210

2 Department of Physics and Astronomy, The University of Tennessee, Knoxville, Tennessee 37996, USA

3 Materials Science and Technology Division, Oak Ridge National Laboratory, Oak Ridge, Tennessee 37831, USA

4 Department of Physics and Astronomy and Stewart Blusson Quantum Matter Institute, University of British Columbia, Vancouver, B.C., Canada, V6T 1Z1

5 Computer Science & Mathematics Division and Center for Nanophase Materials Sciences, Oak Ridge National Laboratory, Oak Ridge, Tennessee 37831, USA

* Corresponding author edagotto@utk.edu

## Abstract

We introduce a variational state for one-dimensional two-orbital Hubbard models that intuitively explains the recent computational discovery of pairing in these systems when hole doped. Our Ansatz is an optimized linear superposition of Affleck-Kennedy-Lieb-Tasaki valence bond states, rendering the combination a valence bond liquid dubbed Orbital Resonant Valence Bond. We show that the undoped (one electron/orbital) quantum state of two sites coupled into a global spin singlet is exactly written employing only spin-1/2 singlets linking orbitals at nearest-neighbor sites. Generalizing to longer chains defines our variational state visualized geometrically expressing our chain as a two-leg ladder, with one orbital per leg. As in Anderson's resonating valence-bond state, our undoped variational state contains preformed singlet pairs that via doping become mobile leading to superconductivity. Doped real materials with one-dimensional substructures, two near-degenerate orbitals, and intermediate Hubbard $U/W$ strengths -- $W$ the carrier's bandwidth -- could realize spin-singlet pairing if on-site anisotropies are small. If these anisotropies are robust, spin-triplet pairing emerges.

## Introduction

Quantum Materials merge topological concepts [1], as in Haldane chains with non-local order parameters [2,3], with electronic correlation effects, as in iron-based superconductors with robust Hubbard $U$ and Hund $J_H$ couplings [4-6]. The Haldane chain started the field of topological materials and some physical realizations are CsNiCl$_3$ [7], AgVP$_2$S$_6$ [8], NENP [9], and Y$_2$BaNiO$_5$ [10]. These chains have a spin gap and protected edge states for open boundary conditions (OBC) [3,11,12]. In iron- and copper-based superconductors, most efforts employ planar geometries. However, Cu-oxide two-leg ladders were widely studied when they were predicted and confirmed to have a spin gap and superconduct [13-19]. Recently, analogous developments occurred in iron ladders BaFe$_2$S$_3$ [20-22] and BaFe$_2$Se$_3$ [23-25] that become superconducting with pressure and display complex properties [20-24,26-36]. However, similar efforts in iron chains TlFeSe$_2$ or TlFeS$_2$, are more limited [37-38].

Within Quantum Materials, quasi one-dimensional (1D) ladders and chains are attractive because powerful computational techniques, such as the Density Matrix Renormalization Group (DMRG) [39,40] and Lanczos [41], allow for the study of model Hamiltonians with accuracy. This removes the veil of theoretical uncertainty in higher dimensions that complicates the comparison theory vs experiment. In particular, this 1D avenue may allow for the challenging study of systems where *both* topology and correlations are simultaneously relevant.

In this context, there are few studies of the effects of hole doping and magnetic-based hole pairing on topological interacting systems. Early work in the *t-J* limit (no double occupancy) for doped S=1 chains, indicated a narrow region of pairing, suppressed by competing ferromagnetism [42]. In related work, triplet superconductivity was also analyzed in 1D [43,44]. More recent efforts using multiorbital Hubbard models unveiled robust tendencies to spin-singlet pairing, an exciting result [45,46]. However, these valuable computational efforts did not provide a concrete mechanism as explanation. In particular, we lack a simple intuitive picture connecting the topological properties of Haldane chains and the emergence of hole pairs in Hubbard models. Developing such a simple ``cartoon'' may allow generalizations to other systems and also facilitate the experimental search for realizations in particular materials.

Here we fill this conceptual gap. Our main conclusion is illustrated in Fig. 1. We consider the two-orbital Hubbard model on a chain, Fig. 1(a), using the two orbitals *a* and *b* as legs of a mathematically equivalent two-leg ladder, Fig. 1(b). We rely on a hereby proposed variational state: the orbital generalization of the resonating valence-bond concepts [47]. We employ preformed spin-1/2 singlets as in the original formulation but now in the enlarged space spanned by the real chain in one direction and the orbital index in another, Fig. 1(c). More simply, our state -- the Orbital Resonant Valence Bond (ORVB) -- is an optimized linear combination of Affleck-Kennedy-Lieb-Tasaki (AKLT) valence bond solids [3,48,49], rendering the proposed state a liquid. Doping this state with two holes in principle could break two singlets. But when holes are close to one another, they break only one singlet minimizing the energy and leading to an effective *singlet* hole pairing, Fig. 1(d), in agreement with computational results [45,46]. Our undoped and doped states are variational, not exact, but they capture the essence of the problem, as shown below.

Our conclusions are not obvious: naively there are preformed triplets at each site because of the robust $J_H/U$. Actually, we found that triplet pairing – a rarity [50,51] -- becomes stable when easy-plane anisotropies are not negligible. However, doping particular quasi-1D materials -- with two active fairly equivalent orbitals and weak spin anisotropy -- should lead instead to singlet pairing.

The present effort for two-orbital chains has qualitative implications for other multiorbital systems, such as iron-based superconductors. More specifically, our results, and the hole binding found in multiorbital ladders [33], show that magnetic fluctuations induce pairing in repulsive Hubbard models. In this framework, these efforts are as important as the theoretical studies of Cu ladders in the 1990s [13,15]: if pairing occurs convincingly in 1D systems, the same Hamiltonian may induce analogous tendencies in higher dimensions where many-body techniques are not as accurate.

## Results

**Model.** We use a canonical two-orbital Hubbard model with kinetic energy and interaction terms written as $H = H_K + H_I + H_D$. The tight-binding portion is

$$H_K = \sum_{i\sigma\gamma\gamma'} t^{\gamma\gamma'}(c_{i\gamma\sigma}^\dagger c_{i+1\gamma'\sigma} + H.c.) \tag{1}$$

where $c_{i\gamma\sigma}^\dagger$ ($c_{i\gamma\sigma}$) creates (destroys) an electron at site $i$ of a chain, orbital $\gamma$ ($a$ and $b$ in our case, although our Hamiltonian notation is generic for arbitrary number of orbitals), and spin projection $\sigma$. The nearest-neighbor (NN) electron hopping is here a 2x2 orbital-space unit-matrix, i.e. $t^{\gamma\gamma'} = t\,\delta_{\gamma\gamma'}$, with $t$ the energy unit throughout the publication. The non-interacting bandwidth is $W = 4.0t$. The hopping symmetry between the two orbitals, and the absence of crystal-field splitting, prevents the appearance of the orbital-selective Mott physics recently studied in related multiorbital models [52-54].

The electronic interaction is standard for multiorbital fermionic systems [55]:

$$\begin{aligned} H_I = &\, U\sum_{i\gamma} n_{i\gamma\uparrow} n_{i\gamma\downarrow} + \left(U' - \frac{J_H}{2}\right)\sum_{i\,\gamma<\gamma'} n_{i\gamma} n_{i\gamma'} - \\ &\, 2J_H \sum_{i\,\gamma<\gamma'} \mathbf{S}_{i\gamma} \cdot \mathbf{S}_{i\gamma'} + J_H \sum_{i\,\gamma<\gamma'} (P_{i\gamma}^\dagger P_{i\gamma'} + H.c.) \end{aligned} \tag{2}$$

The first term is the intraorbital Hubbard repulsion $U$. The second contains the interorbital repulsion at different orbitals, with the usual relation $U' = U - 2J_H$ due to rotational invariance. The third term involves the Hund's coupling $J_H$, and the last term represents the on-site interorbital electron-pair hopping ($P_{i\gamma'} = c_{i\gamma'\uparrow} c_{i\gamma'\downarrow}$).

Later it will also be important to incorporate an easy-plane anisotropy component ($D>0$):

$$H_D = D \sum_i (S^z_{ia} + S^z_{ib})^2 \tag{3}$$

The spin 1/2 operators $(S^x, S^y, S^z)$ are defined as $S^\alpha_{i\gamma} = \left(\frac{1}{2}\right) \sum_{\sigma\sigma'} c^\dagger_{i\gamma\sigma} \sigma^\alpha_{\sigma,\sigma'} c_{i\gamma\sigma'}$ via Pauli matrices. For our results we used the Lanczos method as well as DMRG, with up to $m$=1800 states and truncation errors below $10^{-6}$ as in previous investigations [45].

**Undoped two-orbital Hubbard model at intermediate *U/W* vs Haldane state.** We focus on multiorbital models in iron-based superconductors where ladders and chains can be synthesized, but our results are valid for other transition metal compounds. Iron superconductors are ``intermediate'' between weak and strong coupling, and *U/W* ≈ 1 is considered realistic [4-6]. Because the iron family is not at *U/W* >> 1, a pure spin model is not appropriate and interacting itinerant fermions must be used.

Consider first whether the model discussed here -- with mobile electrons, intermediate *U/W*, and hopping unit matrix -- is smoothly connected to the Haldane limit. At one particle per orbital and *U/W* >> 1 -- with concomitant growth of $J_H$ fixed at the often used ratio $J_H/U$=1/4 [5,58] -- our model certainly develops S=1 states at every site, antiferromagnetically Heisenberg coupled. To analyze if intermediate *U/W* ≈ 1 and strong coupling *U/W* >> 1 (with S=1 spins onsite) are qualitatively similar, we compute with DMRG the entanglement spectra ES [59]. For example, at *U/W*=1.6 where hole-binding is maximized (see below), Figs. 2(a,b) indicate that increasing $J_H/U$ the Hubbard ES clearly resembles the S=1 chain ES [60].

However, our model is not merely a S=1 chain: the inset of Fig. 2(c) indicates that the von Neumann entropy [61-63] $S_{VN}$ converges to ln(2) (S=1 chain result) only at *U/W* ≈ 5 and beyond. At typical couplings of iron compounds, $S_{VN}$ is approximately double the *U/W* >> 1 limit. Thus, the two-orbital Hubbard model qualitatively resembles the Haldane chain, but at *U/W* ≈ 1 there are quantitative differences likely caused by non-negligible charge fluctuations.

Consider now the evolution increasing *D/t*. Recent work found a transition between the gapped Haldane region and a gapped state with trivial topology [46]. In Fig. 2(c) indeed $S_{VN}$ at fixed *U/W* = 1.6 and $J_H/U$ =0.25 does not evolve smoothly from *D/t*=0 -- connected to the large *U/W* Haldane limit -- to the anisotropic large *D/t* ``XY'' limit. The ground state in this limit has a spin triplet with zero *z*-projection at every site, and no edge states. At 0.1 < *D/t* < 0.2, an abrupt change occurs and eventually $S_{VN}$ → ln(1) as *D/t* grows, compatible with a product state of zero *z*-projection triplets [see discussion below, Eq.(4)].

In summary, although with quantitative differences, the undoped Hubbard model qualitatively resembles the Haldane chain as long as *D/t* does not cross a threshold beyond which edge states disappear and a topologically trivial regime develops.

**Pairing in the doped two-orbital Hubbard model.** Our main focus is why pairing occurs and why in the channel it occurs. However, before addressing these issues, let us review and extend recent studies about hole-pair formation and pair-pair correlations in the doped two-orbital Hubbard model. This analysis will provide hints for the intuitive explanation. In Fig. 3(a) the 2-holes binding energy vs $U/W$ is shown, parametric with $J_H/U$. This binding energy is defined as $\Delta E = E(2)-E(0) - 2[E(1)-E(0)]$, with E(M) the ground state energy with M holes (zero holes refers to the half-filled state with one electron per orbital). When $\Delta E$ becomes negative, it signals a 2-holes bound state. Clearly, Fig. 3(a) indicates pair formation with maximum $|\Delta E|$ at $1<U/W<2$, as in [45], and growing with increasing $J_H/U$ (note $J_H/U$ should be less than 1/3 to remain smaller than $U'/U$ due to the constraint $U' = U - 2 J_H$). At $U/W \gg 1$, ferromagnetism for 1 and 2 holes --see discussion below -- prevents pairing suggesting that directly doping the S=1 chain is not the proper theoretical approach. In Fig. 3(b) we show new results, now increasing $D/t$ at fixed $J_H/U$ =0.25. Robust pairing is observed again. However, while the binding curves are almost identical at $D/t$=0.0 and 0.2, at larger $D/t$ they rapidly increase in magnitude. This reflects qualitative differences in pairing, compatible with the von Neumann analysis increasing $D/t$ in Fig. 2(c) that indicated a topological change in the same $D/t$ range.

The qualitative transition in Fig. 3(b) also occurs in Figs. 3(c,d) where pairing correlations are shown. At $D/t$=0 and $J_H/U$ = 0.25, i.e. doping a region smoothly connected to the Haldane chain, spin-singlet pairing dominates (triplet is exponentially suppressed). With increasing $D/t$ at fixed hole density $x$, a transition from singlet to triplet dominance is observed. For example, in panel (d) we observe that spin-triplet pairing, heavily suppressed at $D/t$=0, instead dominates as $D/t$ increases (while singlet is exponentially suppressed).

The ``global summary'' is in Fig. 4 based on a large set of DMRG data. It contains a phase diagram varying $D/t$ and doping $x$, with only a few representative points displayed. The red region near $x$=0 and $D/t$=0 is where the model resembles the Haldane state according to the entropy entanglement. Here, at light doping $x$ singlet-pairing dominates at intermediate $U/W$, but eventually other non-superconducting channels (SDW and CDW) take over as $x$ grows. Increasing $D/t$, at small $x$ a transition from singlet- to triplet-dominated pairing occurs .In the singlet regime, holes are primarily located at nearest-neighbor sites and different orbitals, while in the triplet regime they are primarily at the same site in different orbitals.

The DMRG results unveiled a parameter space region (small $D/t$, low hole-doping, intermediate $U/W$, and robust $J_H/U$) where superconducting spin-singlet correlations dominate. These results are surprising. First, the connection with the S=1 chain suggests that antiferromagnetic (AFM) fluctuations are short-range and perhaps not sufficiently strong for pairing. Second, at every site and at intermediate-strong $U/W$ a nonzero magnetic moment develops due to $J_H/U$. Naively, these same-site electrons can be considered as local preformed triplets. Then, after hole doping the resulting ground state could be envisioned as these triplets becoming mobile. For $D/t > 0.2$, this naive perspective is compatible with numerical results in Fig. 4. However, better understanding the on-site spin-triplet pairing will require further work because in principle a Hund coupling $J_H/U=1/4$ is not sufficient to overcome the inter-orbital repulsion. The anisotropy *D*, which influences on the energy, seems important to stabilize the triplet pairing as our computational results indicate. The product state wave function Eq. (4) is a good variational approximation (undoped system), exact as *D* diverges:

$$|TPS\rangle = \prod_i^N |1,0\rangle_i = \prod_i^N \frac{1}{\sqrt{2}}(|\uparrow_{ia},\downarrow_{ib}\rangle + |\downarrow_{ia},\uparrow_{ib}\rangle) \qquad (4)$$

Then, why singlets dominate at small *D/t*? Although in a Haldane regime all triplet correlations must decay exponentially, such reasoning does not explain why singlet pairing is enhanced. Hints for the variational state presented below arise from the AKLT exact solution [3,48,49], where a S=1 spin model was considered employing two auxiliary idealized S=1/2 degrees of freedom at every site. These auxiliary states form spin singlets with other S=1/2 auxiliary states at the next site. Below, we show that the two-orbital Hubbard model shares properties similar to this intuitive idea.

**Variational state for the undoped two-orbital Hubbard chain.** We now introduce a variational state for both the undoped and lightly doped two-orbital Hubbard chain, at intermediate and strong *U/W*. We argue that these ground states can be qualitatively described in terms of S=1/2 spin singlets involving nearest-neighbor (NN) sites, connecting the same or different orbitals. *Our main result is that the small D/t region of the two-orbital Hubbard model has hidden ``preformed'' singlets that become mobile with doping*. Knowing what type of Hubbard model develops pairing, and with what type of hoppings, allow us to predict what characteristics a material must display to realize this physics.

We propose a variational state inspired by an exact equality. Consider first only 2 sites, say 1 and 2, and construct the quantum global spin-zero state using one electron per orbital. With only one orbital this has the canonical expression $|\text{Singlet}_{S=1/2\ 2\text{-sites}}\rangle = \frac{1}{\sqrt{2}}(|1/2,1/2\rangle_1|1/2,-1/2\rangle_2 - |1/2,-1/2\rangle_1|1/2,1/2\rangle_2) = \frac{1}{\sqrt{2}}(|\uparrow_1\downarrow_2\rangle - |\downarrow_1\uparrow_2\rangle)$ where $|1/2,1/2\rangle$ means total spin 1/2 and z-projection ↑, etc.

For two spins 1, the global spin zero state is still relatively simple

$$|\text{Singlet}_{S=1\ 2\text{-sites}}\rangle = \frac{1}{\sqrt{3}}(|1,1\rangle_1|1,-1\rangle_2 + |1,-1\rangle_1|1,1\rangle_2 - |1,0\rangle_1|1,0\rangle_2) \qquad (5)$$

Because in our case each site S=1 arises from two real S=1/2 electrons at each orbital and same site, we use the $|1,1\rangle_1 = |\uparrow_{1a}\uparrow_{1b}\rangle$, $|1,-1\rangle_1 = |\downarrow_{1a}\downarrow_{1b}\rangle$, $|1,0\rangle_1 = (1/\sqrt{2})(|\uparrow_{1a}\downarrow_{1b}\rangle + |\downarrow_{1a}\uparrow_{1b}\rangle)$ notation and an analogous expression at site 2. Then, simple algebra leads to

$$|\text{Singlet}_{S=1\ 2\text{-sites}}\rangle = \frac{1}{\sqrt{3}}(|\uparrow_{1a},\uparrow_{1b},\downarrow_{2a},\downarrow_{2b}\rangle + |\downarrow_{1a},\downarrow_{1b},\uparrow_{2a},\uparrow_{2b}\rangle) -$$
$$\frac{1}{2\sqrt{3}}(|\downarrow_{1a},\uparrow_{1b},\uparrow_{2a},\downarrow_{2b}\rangle + |\uparrow_{1a},\downarrow_{1b},\downarrow_{2a},\uparrow_{2b}\rangle + |\uparrow_{1a},\downarrow_{1b},\uparrow_{2a},\downarrow_{2b}\rangle + |\downarrow_{1a},\uparrow_{1b},\downarrow_{2a},\uparrow_{2b}\rangle) \quad (6)$$

What is remarkable is that this last expression can be exactly rewritten as a combination of S=1/2 singlets, involving either different or the same orbitals:

$$|\text{Singlet}_{S=1\ 2\text{-sites}}\rangle = \frac{-1}{\sqrt{3}}[\frac{1}{\sqrt{2}}|\uparrow_{1a},\downarrow_{2b} - \downarrow_{1a},\uparrow_{2b}\rangle \frac{1}{\sqrt{2}}|\uparrow_{1b},\downarrow_{2a} - \downarrow_{1b},\uparrow_{2a}\rangle$$
$$-\frac{1}{\sqrt{2}}|\uparrow_{1a},\downarrow_{2a} - \downarrow_{1a},\uparrow_{2a}\rangle \frac{1}{\sqrt{2}}|\uparrow_{1b},\downarrow_{2b} - \downarrow_{1b},\uparrow_{2b}\rangle] \quad (7)$$

Intuition is gained when this result is represented visually, Fig. 5, where we have rewritten exactly Eq. (7) doubling the number of valence-bonds states for an easier extrapolation to more sites (using the total spin at each site $\mathbf{S}_i = \mathbf{S}_{ia} + \mathbf{S}_{ib}$, the identity $(\mathbf{S}_{ia} + \mathbf{S}_{ib}) \cdot (\mathbf{S}_{ia} + \mathbf{S}_{ib}) = \mathbf{S}_{ia} \cdot \mathbf{S}_{ia} + \mathbf{S}_{ib} \cdot \mathbf{S}_{ib} + \mathbf{S}_{ia} \cdot \mathbf{S}_{ib} + \mathbf{S}_{ib} \cdot \mathbf{S}_{ia}$ also helps in this context).

Our exact result is counterintuitive: with perfect S=1 states at each site, the total spin-zero state of the two-orbital two-sites Hubbard model can be represented exactly as a linear combination of S=1/2 singlets. This resembles the original AKLT perspective [3] although here applied to a fermionic system. Figure 5 is as in the views of Anderson and Affleck et al., but with orbitals as legs of a two-leg ladder, with these legs only connected by $J_H$ (no inter-leg hopping).

Then, intuitively, as in the AKLT states, the undoped state has preformed S=1/2 singlet pairs in all possible arrangements that upon doping should become mobile, leading to spin-singlet pairing dominance. Thus, we predict that doping real quasi-one-dimensional materials with two dominant nearly-degenerate orbitals should lead to superconductivity in the spin-singlet channel if anisotropies are not large. We need two ``similar'' orbitals because we used a 2x2 unit hopping matrix.

How is this generalized to more sites? The two-site exact result in Fig. 5 assuming PBC establishes a rule: at each elementary 2x2 plaquette only one singlet can be used, either along a diagonal or along a leg. Each ladder site S=1/2 can be used only once: after forming a singlet they disappear from the picture. The two-site example has other properties common to a longer chain. The global singlet state is *even* under the exchange of orbitals *a* and *b* and also even under a reflection with respect to the middle of the plaquette. Now extending to more sites becomes natural. For example, in Fig. 6 we show the 16 valence bond states needed for 4 sites using PBC, as well as the ``representative'' of each class (i.e. applying to a representative translations and orbital exchange, the full original class can be reconstructed). We remark again that, by construction, all states have perfect S=1 spins at every site, as in the AKLT setup, even using spin-1/2 singlets as building blocks.

**Lanczos overlaps.** How accurate is this state? Using Lanczos, we calculated the normalized exact ground state GS of the two-orbital Hubbard model in short chains and computed the overlap with the ORVB linear combination of the individual AKLT-like states of Fig. 6. The coefficients for each class were optimized to maximize the global overlap, arriving to a final normalized-to-one state dubbed ORVB. Care must be taken because the individual AKLT components do not form an orthogonal set. By this procedure, at $U/W$=20 and 2 sites, $|\langle ORVB|GS\rangle|$ is virtually 1, because double occupancy is much suppressed, as in Eq. (7). Using 4 sites, the binding energy is now optimized at $U/W$=4. At this coupling and size, $|\langle ORVB|GS\rangle|$ = 0.95 indicating that ORVB state is a good representation of the ground state.

We extended to more sites using PBC. As already explained, below when ``classes'' are mentioned for the undoped case, they represent groups of valence bond states related by applying translations and orbital exchange to a particular representative. For 6 sites, there are 8 classes and at $U/W$=2, where binding is maximized, the overlap is 0.79. As the lattice size grows, other configurations involving longer S=1/2 singlets and especially doubly occupied orbitals will contribute to the ground state because the optimal $U/W$ where binding is maximized is reduced towards the intermediate range. But finding a robust 0.79 overlap with 6 sites indicates that ORVB is a good variational state. The same occurs for 8 sites: here the number of classes is 16, and when $U/W$=1.5 is chosen because it optimizes the binding energy, the overlap $|\langle ORVB|GS\rangle|$ remains robust at 0.61. Increasing the system size, the optimal binding converges to intermediate $U/W$, see Fig. 3.

Should we worry about a reducing overlap with increasing size? As example consider the simple $(\pi,\pi)$ spin staggered state $|\text{stagg}\rangle = |\uparrow\downarrow\uparrow\downarrow...\rangle$ of the S=1/2 Heisenberg model two-dimensional square lattice. For a 2x2 cluster its overlap with the true ground state is 0.58 but for the 4x4 cluster it decreases to 0.29. However, $|\text{stagg}\rangle$ is certainly a good variational state. A reducing overlap is natural because with increasing cluster size the fraction of the total Hilbert space spanned by simplified variational states -- such as proposed here for two-orbitals or the spin staggered state for Heisenberg models -- rapidly decreases. Thus, after confirming the overlap is robust for small clusters, what matters more is whether the proposed state captures the essence of the ground state, as shown next.

**Doped variational state and superconductivity.** Let us generalize our variational state to the doped case, a topic barely addressed in the AKLT context. In our DMRG studies in Fig. 3 and in previous efforts [45], we found that in the 2-holes ground state the largest-weight configuration occurs when holes are placed at NN sites and in different orbitals. Having the two holes in the same leg is not optimal because they collide: with one hole per leg they can move without obstacles, while taking advantage of the effective attraction in the variational state arrangement. For this reason, and to reduce complexity, our proposed doped state will have only one hole per orbital and will be obtained primarily from the undoped state by *removing one diagonal singlet*.

This is exemplified for 4-sites PBC in Fig. 7 (left) where classes are shown. Note that the diagonal character of the 2x2 hopping matrix establishes that the number of holes per orbital is conserved.

Additional remarks: (1) here we use the unit hopping matrix, but in most realistic situations a nonzero crystal field among the orbitals (rendering them non-equivalent even after a change of basis) as well as inter-orbital hoppings and non-equal diagonal hoppings could be present. What occurs in these conditions remains to be studied. (2) While at very large $U/W$ the AKLT guidance should work well, at intermediate $U/W$ the form of the orbital hopping matrix influences on the energy. Then, in the ORVB Ansatz, the states with all inter-orbital singlets will not have the same weight as the states with all intra-orbital singlets.

Two extra ingredients are needed. First, quantum mechanically each hole in each orbital is ``oscillating'' (zero-point motion) via the intraorbital hopping because these are not frozen holes. Then, the configuration with 2 holes in the same rung must be included because it is generated by oscillations within the bound state. Second, to avoid unpaired S=1/2 electrons left and right of that 2-holes rung, a singlet across is required, as in the three classes in Fig. 7 (right). Indeed a $\pi$-shift across-the-hole develops in the spin correlations of the two-orbital Hubbard ground state [45]. This also occurs in one-orbital $t$-$J$ models [64-65]. For completeness, singlets across-the-hole were also added for diagonal hole configurations as in the bottom left class of Fig. 7.

This procedure resembles qualitatively the exact solution of the infinite $U$ single-orbital Hubbard chain [66]: holes and spins are independent in this limit. Our mobile holes can be visualized as *effectively inserted in between the original singlets of the undoped valence bond state*.

The ORVB state generalized from Fig. 7 but now for 6 sites and 2 holes requires 23 classes [for 2 holes, to generate all states not only translational symmetry and exchange of orbitals are needed, but also reflection (parity) with respect to the middle]. The overlap with the 2-holes Lanczos exact GS at $U/W$=2 is 0.59. For 8 sites and 2 holes, 84 classes are needed, and the overlap at $U/W$=1.5 is 0.48. These are good numbers, but more important is how *qualitatively* these states capture the essence of the problem. For instance, from the 6-sites exact GS the spin-spin correlations can be measured using special projection operators [45,67] for when the mobile holes are at their highest-probability ground state position [33,45]. The Lanczos and our variational results are contrasted in Figs. 8(a,b). The agreement is remarkable. While away from the holes the pattern resembles the undoped case, with ferromagnetic (FM) rungs and AFM legs [45], near the hole the AFM correlation ``across-the-hole'', typical of carriers in an AFM background, is reproduced. Moreover, a *puzzling* FM link diagonally placed opposite to the diagonal of holes is also observed. Naively this may suggest that triplets are needed in the undoped variational state along diagonals. However, this effective FM correlation is merely a consequence of the mobility of the holes that displace the original on-site FM triplet -- contained in our variational state by construction because of the AKLT projection to spin 1 at every site -- from rung to diagonal.

Panel Fig. 8(c) also shows that at very large $U/W$ our proposed picture breaks down. In this regime, the effective AFM superexchange weakens because it scales as $1/U$. As a consequence, alternative tendencies such as ``double-exchange'' as in the manganite context [55-57] are enhanced leading to ferromagnetism to improve the kinetic energy of the now unbounded holes. This suggests that simply doping the S=1 Haldane chain may not be sufficient, but $U/W$ must be limited to intermediate values to avoid the ferromagnetic competition. However, in the reported hole-binding range the total spin quantum numbers are qualitatively compatible with those of our variational picture. This optimal values of $U/W$ are also compatible with our ladder work [33] and with alternative explanations focused on pairing amplitudes and effective exchange interaction optimizations [68,69].

**Varying $D/t$.** When easy-plane anisotropies are included, the spin-triplets product state at each site TPS, Eq.(4), becomes increasingly a better approximation as $D/t$ grows (at $D/t$ diverges, TPS is the exact ground state). This evolution is illustrated in Fig. 9. In (a), the half-filling overlaps $|\langle ORVB|GS\rangle|$ and $|\langle TPS|GS\rangle|$ are shown. The ORVB (TPS) overlap decreases (increases) with increasing $D/t$, as expected. Note that the TPS state used here is crude: larger overlaps in the $D/t$ range shown could be obtained adding fluctuations but this is irrelevant for our main focus i.e. the origin of the spin-singlet pairing at small $D/t$.

In addition, we observed that the spin-spin correlations involving the $z$ components within the two orbitals of the same rung correctly evolve with increasing $D/t$. At small $D/t$, they are FM because a spin close to S=1 forms at each site. At large $D/t$, they become AFM because only the spin zero $z$-projection component survives in the XY product state. Also note the ORVB and TPS states states are not orthogonal to one another, but their overlap is very small (0.02 for $N$=8, 0.07 for $N$=6, both at $D$=0 PBC).

The case of 2 holes, panel (b), is more interesting. Here a *level crossing* occurs [panel (c)]: at small $D/t$ the ORVB 2-hole-doped state has a nonzero overlap with the 2-hole Lanczos state because both states have quantum number (-1) under orbital exchange (orbital antisymmetric). However, at $D/t \approx 0.125$, where the von Neumann entropy Fig. 2(c) signaled a qualitative transition in the undoped case, a level crossing occurs in the Lanczos ground state. At large $D/t$ the quantum number under orbital exchange becomes (+1) leading to a nonzero overlap with the orbital symmetric 2-hole TPS state.

## Discussion

We introduced a variational state for the undoped and hole-doped two-orbital Hubbard chain, verified its accuracy, and explained the development of spin-singlet pairing upon doping. Our analysis relies on valence bond states defined in an extended ladder-like geometry spanned by the real chain in the long axis and the number of orbitals in the short axis. Our variational state is an optimized linear combination of AKLT singlets. Using DMRG and Lanczos, we find excellent agreement with our variational predictions at intermediate $U/W$. The entanglement spectra and von Neuman entropy indicate that the undoped intermediate $U/W$ regime is connected to the Haldane limit at $U/W \gg 1$. However, in the realm of spin models at $U/W \gg 1$, a strong competition with ferromagnetism upon doping prevents pairing from occurring.

Our variational state relies on a mathematical expression involving two S=1 spins at NN sites, with one electron per orbital. The global spin-zero state of these two sites is written exactly exclusively using spin-1/2 singlets linking electrons at those NN sites, involving the same or different orbitals. When extended to more sites, the proposed variational state is an optimized linear combination of spin-1/2 singlets in all possible NN arrangements. Note that same-rung singlets are excluded because of the large ferromagnetic Hund coupling. Using Lanczos, our variational state was shown to be a good approximation to the true ground state for small easy-plane anisotropy $D/t$, robust Hund coupling $J_H/U$, intermediate $U/W$, and light hole doping.

The preformed NN spin singlets -- Cooper pairs -- become mobile upon hole doping, and will form a coherent superconducting state if employing the canonical BCS-like product-state construction, at least within the limitations of one dimensionality that only allow for power-law decays. A weak coupling among chains will render the state truly superconducting with long-range order, as in two-leg Cu- or Fe-based ladders. Because of the small size of the Cooper pairs, the coherent state is likely in the Bose Einstein condensation class.

What occurs if more orbitals are used? The pioneering work of Haldane established that integer and half-integer spin chains are intrinsically different. Thus, if we use three orbitals and still a unit-matrix hopping, a generalized ORVB state with a spin gap due to the presence of spin singlets should not be a good variational state. However, using four orbitals we should return to the class of two. The situation becomes more complicated, and difficult to predict, if in addition to modifying the number of orbitals we also add crystal fields, inter-orbital hoppings, or assign different values in the diagonal for different orbitals. In this case, the subject is totally open. First indications [33] suggest that binding is possible for ladders with a non-trivial hopping matrix but probably the undoped state is not topological as in the Haldane chain. These many open issues will be addressed in the future.

In summary, our study combines topological concepts with correlation effects. Pairing emerges with hole doping via the liberation of preformed spin-1/2 singlets already contained in the undoped limit. For experimentalists to realize our model the challenge is to find quasi-1D materials with two nearly-degenerate dominant active orbitals, and with similar overlaps along the chain so that the hopping matrix is nearly the unit matrix, as in our calculations. How robust the Haldane regime is with regards to small deviations from this hopping symmetric case, as well as the introduction of an orbital small crystal-field splitting, remains to be investigated (the orbital-selective Mott phase is close in parameter space [52-54]). Doping the existing physical realizations of undoped Haldane chains [7,10] can provide a starting point towards the predicted superconductivity, but, again, intermediate $U/W$ is a more attractive parameter region than $U/W \gg 1$. For this reason the iron-superconductors family provides a natural starting point, although realizations could also be found in other multiorbital-active compounds.

## Methods

**Operators.** To address pairing, we defined a general pair creation operator as

$$\Delta^{\gamma\gamma'}_{(i,j)\pm}{}^{\dagger} = \frac{1}{\sqrt{2}}(c^{\dagger}_{i\gamma\uparrow}c^{\dagger}_{j\gamma'\downarrow} \pm c^{\dagger}_{i\gamma\downarrow}c^{\dagger}_{j\gamma'\uparrow}) \tag{8}$$

where $i,j$ are sites, $\gamma,\gamma'$ are orbitals ($a$ or $b$), and ± sign represents a spin singlet or triplet. We only focused on two different pair operators: (1) nearest neighbor pair that is odd under spin (singlet) and under orbital exchange ($S^{\dagger}_{nn}(i)$ below), (2) on-site inter-orbital pair that is even under spin (triplet) and under orbital exchange ($T^{\dagger}_{on}(i)$ below). These are defined as

$$\Delta_S = S^{\dagger}_{nn}(i) = \Delta^{ab}_{(i,i+1)-}{}^{\dagger} - \Delta^{ba}_{(i,i+1)-}{}^{\dagger}$$
$$\Delta_T = T^{\dagger}_{on}(i) = \Delta^{ab}_{(i,i)+}{}^{\dagger} \tag{9}$$

In the main text, we often refer to $S^{\dagger}_{nn}(i)$ as odd diagonal singlet and to $T^{\dagger}_{on}(i)$ as on-site triplet, where ``diagonal'' and ``rung'' refers to the ladder representation of a two-orbital chain Fig. 1. Using these pair creation operators, we study the decay of the pair-pair correlations

$$P_S(R) = \frac{1}{N_R}\sum_i \langle S^{\dagger}_{nn}(i)S^{\dagger}_{nn}(i+R)\rangle$$
$$P_T(R) = \frac{1}{N_R}\sum_i \langle T^{\dagger}_{on}(i)T^{\dagger}_{on}(i+R)\rangle \tag{10}$$

where $N_R$ represents the number of total neighbors at distance $R$ with respect to site $i$, summed over all sites.

## Code and Data Availability

The computer codes used in this study are available at https://g1257.github.io/dmrgPlusPlus/. The data that support the findings of this study are available from the corresponding author upon request.

## Acknowledgments

We thank A. Moreo and I. Affleck for useful discussions. N. D. P., N. K., A. N., and E. D. were supported by the U.S. Department of Energy (DOE), Office of Science, Basic Energy Sciences (BES), Materials Science and Engineering Division. G. A. was supported by the Scientific Discovery through Advanced Computing (SciDAC) program funded by the U.S. Department of Energy, Office of Science, Advanced Scientific Computing Research and Basic Energy Sciences, Division of Materials Sciences and Engineering. Part of this work was conducted at the Center for Nanophase Materials Sciences, sponsored by the Scientific User Facilities Division (SUFD), BES, DOE, under contract with UT-Battelle.

## Competing interests

The authors declare no competing interests.

## Author contributions

N. D. Patel and E. Dagotto planned the project and provided insight into the understanding of the spin singlet pairing mechanism. N. D. Patel and N. Kaushal performed all the calculations for the multi-orbital Hubbard model. A. Nocera and G. Alvarez developed the DMRG++ computer program.

**Figure Legends**

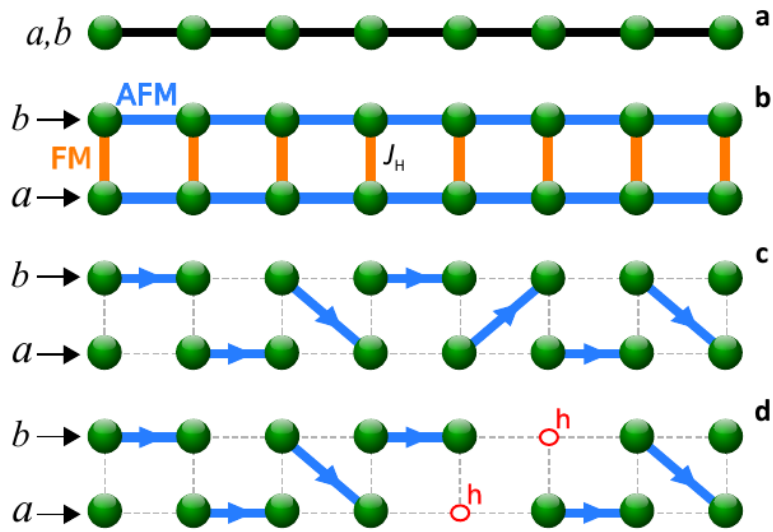

**Fig.1: Summary Main Results**

**a** Sketch of a chain with two active orbitals *a* and *b*. **b** Representation of panel (a) splitting the orbitals into legs forming a fictitious two-leg ladder, with legs only connected by the Hund coupling $J_H$. **c** One component of the variational state proposed in the text. Arrows indicate spin-1/2 singlets linking nearest-neighbor sites. Although their spin is zero, they are oriented objects because singlets are antisymmetric under the exchange of spins. The full ORVB state is an optimized linear combination of all possible arrangements of these singlets i.e. a linear combination of AKLT valence bond solids. **d** Doped state: holes ``h'' are effectively paired when a spin singlet is removed.

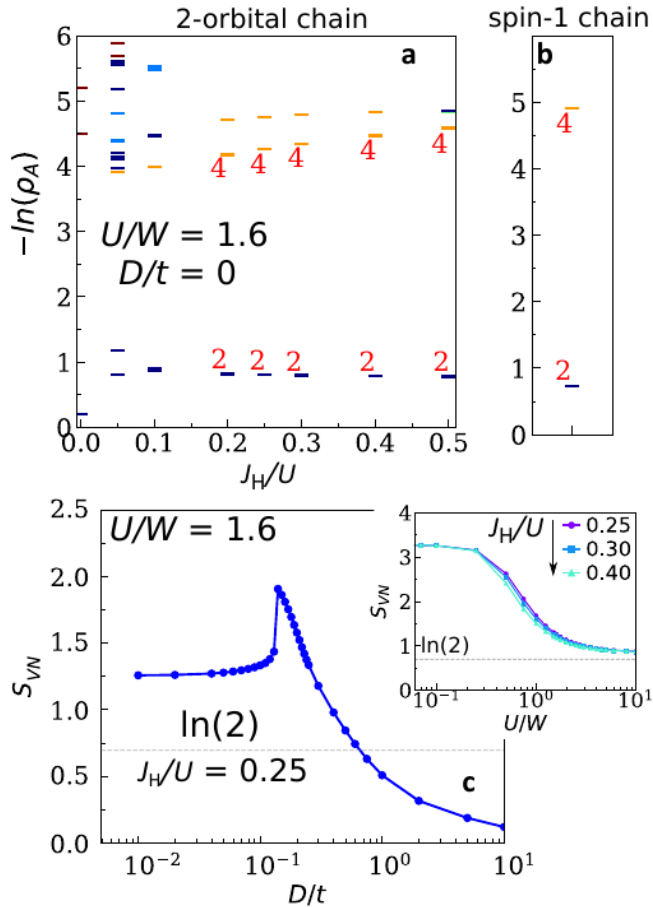

**Fig.2: Entanglement Spectra**

**a** the undoped two-orbital Hubbard chain vs $J_H/U$, at $U/W$ = 1.6. **b** the S=1 Heisenberg chain (both at $D/t$=0). At robust $J_H/U$ in **a**, a two-fold degeneracy is clear in both cases. **c** Von Neumann entanglement entropy ($S_{VN}$) for the undoped two orbital-chain model vs $D/t$, at $U/W$ = 1.6 and $J_H/U$ =0.25. Inset: $S_{VN}$ vs $U/W$ for $D/t$=0 and various $J_H/U$s showing convergence to the $\ln(2)$ of the S=1 chain at $U/W \gg 1$. The DMRG results in (**a-c**) use OBC N = 100 sites, both for Hubbard and Heisenberg S=1 models.

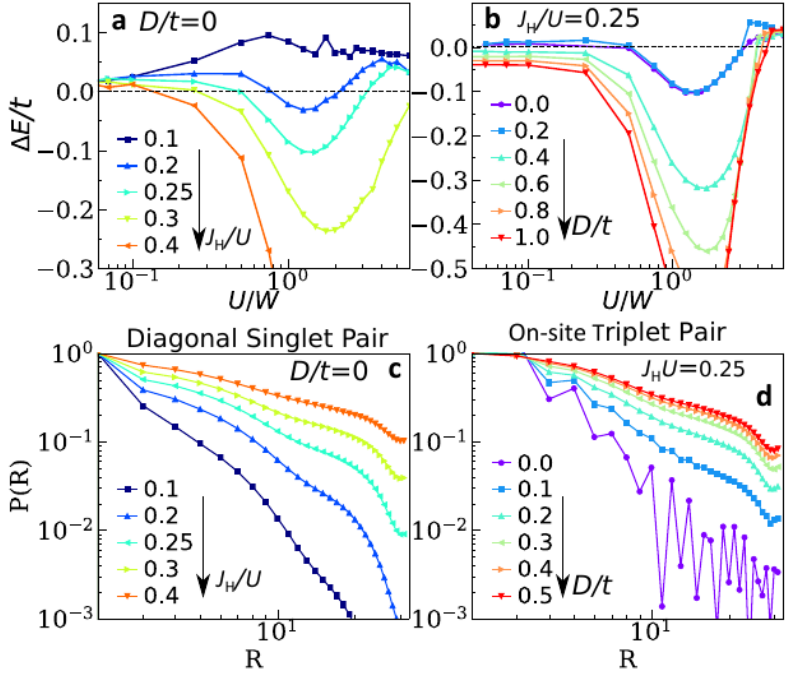

**Fig.3: Binding and Pairing**

Binding energy $\Delta E/t$ vs $U/W$ for various values of **a** $J_H/U$, at $D/t = 0$, and **b** $D/t$, at $J_H/U = 0.25$. In (**a,b**), a 16-sites OBC chain was used and DMRG. **c** Spin-singlet $\Delta_S$ real-space pair-pair correlations P(R) vs distance R, varying $J_H/U$, at fixed $U/W = 1.6$ and $D/t=0$. $\Delta_S$ involves nearest-neighbor sites and different orbitals. **d** Same as **c** but using the on-site triplet operator $\Delta_T$, varying $D/t$, at fixed $U/W = 1.6$ and $J_H/U = 0.25$. In (**c,d**), a 48-sites OBC chain was used and DMRG, neglecting 8 sites at each end to avoid edge effects. Correlations are normalized to the result at distance 2, P(2), to better focus on the large R behavior. For the definition of $\Delta_S$ and $\Delta_T$, see Methods. At **c** the hole doping is $x$=0.042 corresponding to 4 holes, while at **d** $x$=0.083 corresponding to 8 holes. $x$ is the number of holes divided by 96 (48 sites, 2 orbitals). For completeness, we repeated several $\Delta E/t$ calculations in (**a,b**) removing the electron-pair hopping term from Eq.(2) to avoid the impression that this term, arising from Coulomb energy matrix levels [55], may cause the binding. In all cases studied, the binding curves were barely affected by removing the electron-pair hopping.

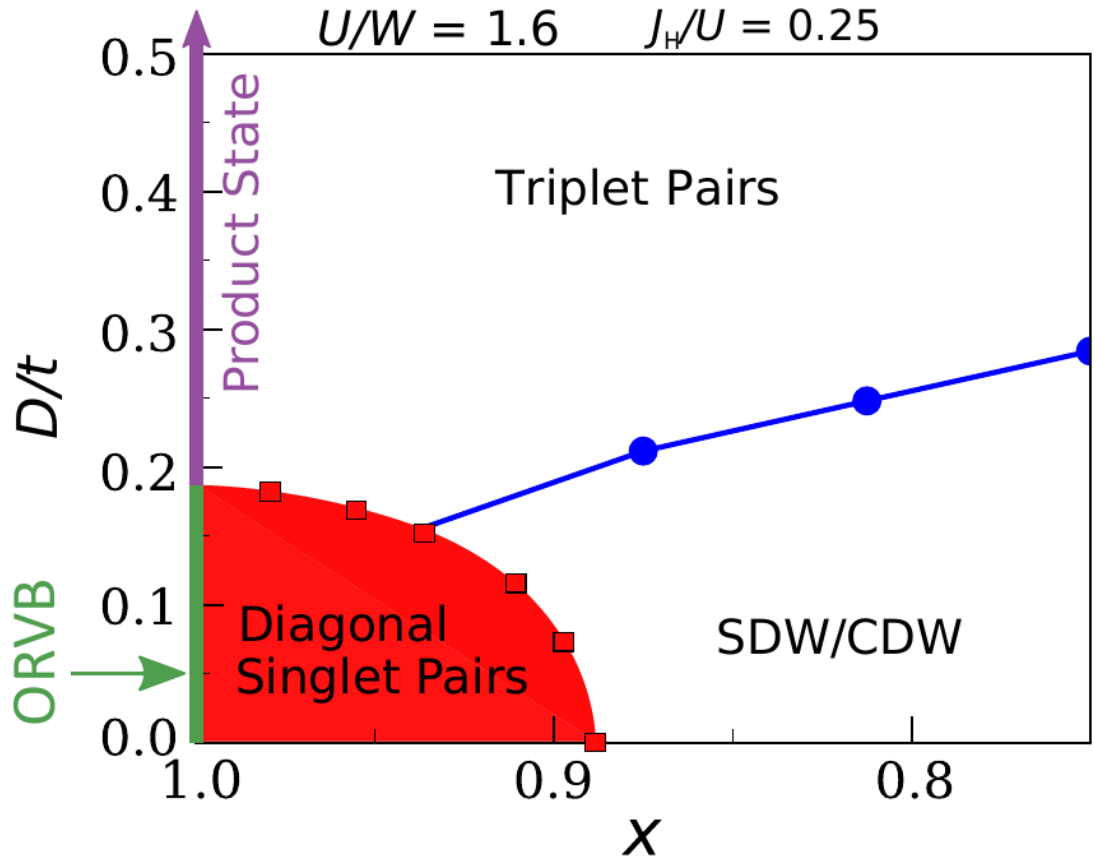

**Fig.4: Phase Diagram**

Qualitative phase diagram varying the easy-plane anisotropy $D/t$ and hole doping $x$, at fixed $U/W$=1.6 and $J_H/U$ = 0.25, using DMRG. Hole density $x$ = 0 represents half-filling where at $D/t$ smaller than ≈ 0.2 the ground state is qualitatively connected to the S=1 Haldane phase. The label ORVB refers to the variational state introduced later in the text. At larger $D/t$, a product state of on-site triplets with zero spin projection Eq. (4) is a good representation of the ground state. Upon doping at small $D/t$, first singlet pairing dominates until at $x$ ≈ 0.1 the spin/charge density wave (SDW/CDW) correlations become stronger. Doping of the product state at $D/t$ ≈ 0.2 or larger leads to spin-triplet pairing over a broad range of doping. We used DMRG and $N$=48 OBC chains to construct the phase diagram. Only a few points are shown with dots, but a denser grid $(x,D/t)$ was analyzed via DMRG.

$$|\text{ORVB}\rangle = \frac{1}{2\sqrt{3}} \left[ \text{(diagrams)} + \text{(diagrams)} - \text{(diagrams)} - \text{(diagrams)} \right]$$

2-sites
2-orbitals
Spin Singlet

**Fig.5: Two sites variational**

Normalized-to-one two-site two-orbital undoped ground state of the Hubbard model at large $U/W$, Eq.(7), when double occupancy is neglected. Shown is the total spin zero state of two sites, with two electrons per site, represented exactly in terms of antisymmetric S=1/2 singlets (blue arrows). *a* and *b* are the two orbitals and (1,2) are the sites. Mathematically this is simply a linear combination of AKLT states. Results are depicted as with periodic boundary conditions (PBC), providing a natural extension beyond two sites, leading to our proposed variational state below.

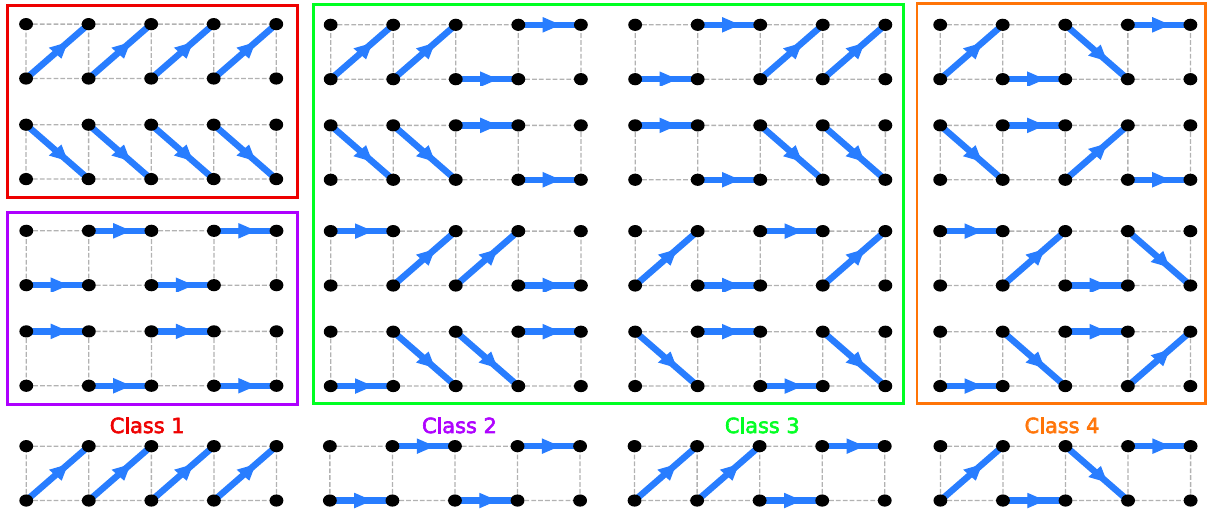

**Fig.6: Four sites variational**

Individual states used for the four-site half-filled two-orbital Hubbard model with PBC. The color-framed states represent 4 distinct ``classes''. A representative of each class is shown at the bottom with the same color convention. By applying translation and orbital exchange for each representative (reflection is not needed for the undoped case), we recover all the states in the upper frames. Note that the 16 AKLT states displayed are not orthogonal to one another. Our proposed variational state is an energy-optimized linear combination of these 16 states (i.e. the weight of the 4 classes is different).

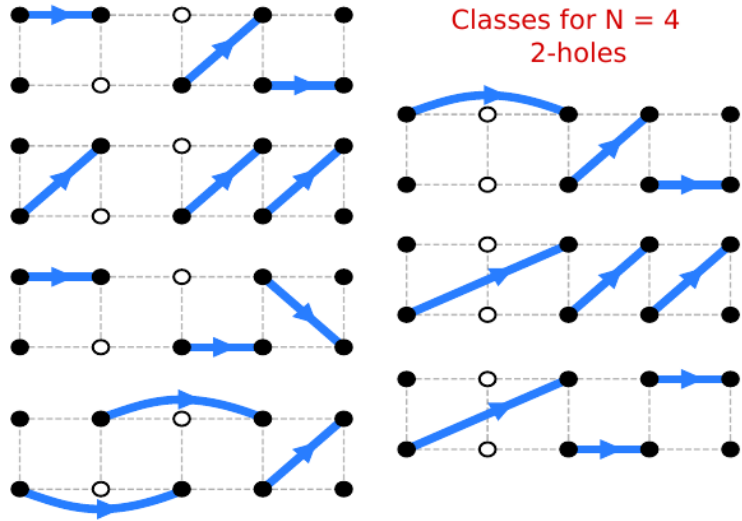

**Fig.7: Doped variational**

A representative of each of the 7 valence bond classes used for a 4-sites chain with PBC and 2-holes doping. Each class state represents a linear combination involving translated, orbitals *a* and *b* exchanged, and parity-inverted states.

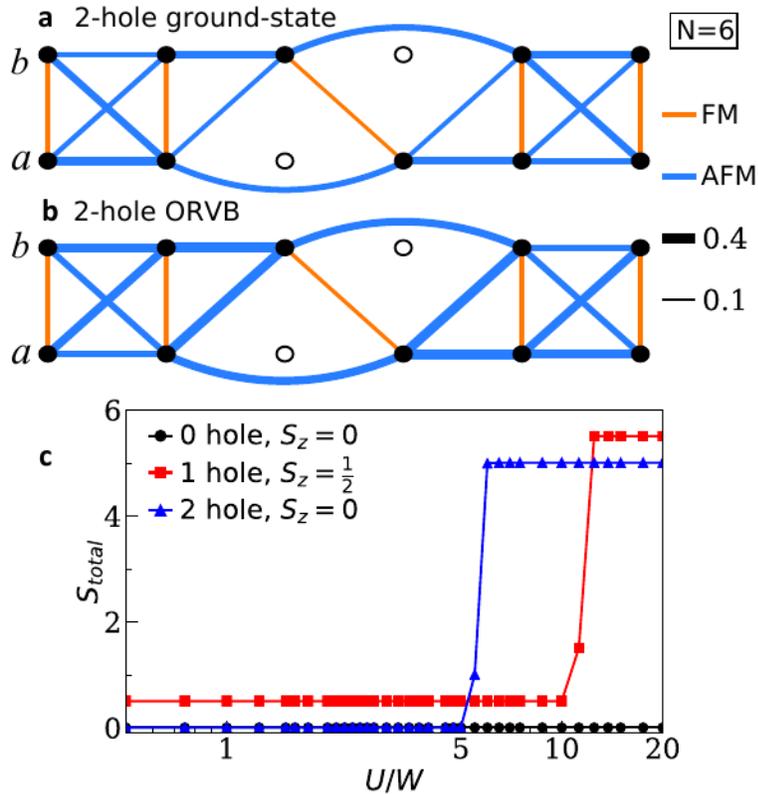

**Fig.8: Variational vs exact**

Schematic of the real-space spin-spin correlations in the two-orbital chain for: **a** the 2-holes exact ground-state and **b** the 2-holes ORVB variational state. The lower (upper) chain represents the orbital $a$ ($b$). The holes, which are of course mobile, are projected to their most likely position in the state via projector operators [45,67], and then spin-spin correlations are measured. Blue (orange) lines represent AFM (FM) bonds with line-thickness proportional to the magnitude of spin correlations. **c** Total ground-state spin quantum number vs $U/W$, for 0, 1, and 2 holes. Calculations (**a,b**) are performed using Lanczos on a 6-sites PBC chain, at $U/W=2.0$, $J_H/U = 0.25$, and $D/t=0$. Here the probability of single occupancy of one orbital is 96% indicating that local moments $S \approx 1$ are well formed. However, the Heisenberg limit is only reached at $U/W>5$ (Fig. 2(c) inset). This suggests that other terms in the strong coupling expansion are of relevance in the regime of binding.

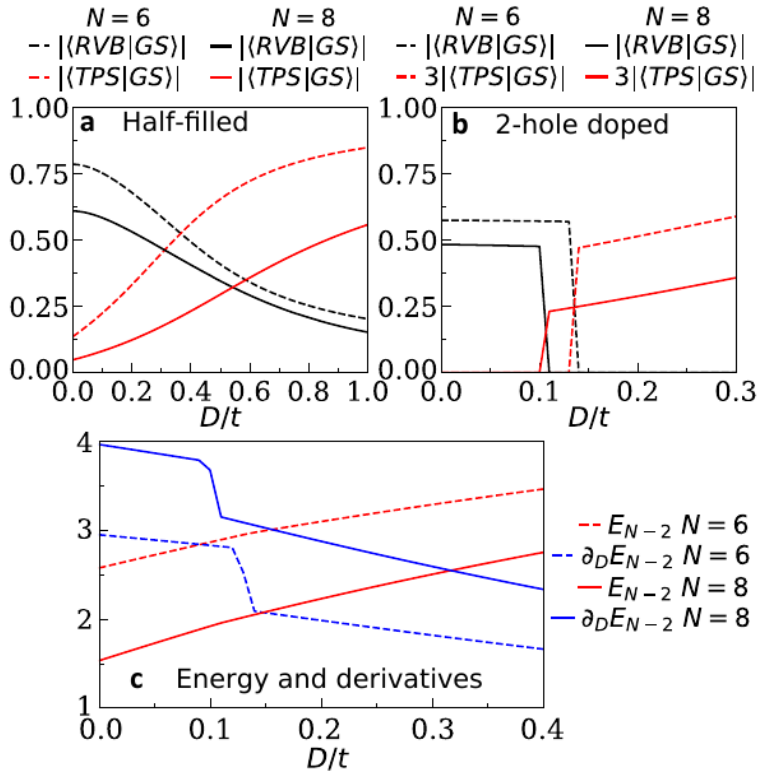

**Fig.9: Overlaps**

Overlaps $|\langle ORVB|GS\rangle|$ and $|\langle TPS|GS\rangle|$ vs $D/t$ for: **a** zero hole (half-filling) and **b** two holes. The discontinuity in **b** indicates a first-order transition due to a level crossing and associated change in the ground state quantum numbers under orbital exchange; **c** ground-state chain energies, as well as their derivatives to emphasize sudden slope changes, for two holes $E_{N-2}$. All results (**a-c**) obtained with Lanczos using $N$=6 and 8 sites and PBC.